\documentclass[a4paper,groupedaddress,showkeys,showpacs,twocolumn]{revtex4}
\usepackage{graphicx}
\usepackage{hypernat}
\usepackage{hyperref}

\begin{document}

\title{Compact in-place gate valve for molecular beam experiments}%
\author{Jochen K\"upper}%
\email[{Author to whom correspondence should be addressed.
   Email:~}]{jochen@fhi-berlin.mpg.de}%
\author{Henrik Haak}%
\author{Kirstin Wohlfart}%
\author{Gerard Meijer}%
\affiliation{Fritz-Haber-Institut der Max-Planck-Gesellschaft, 14195 Berlin,
   Germany}%
\date{\today}%
\begin{abstract}\noindent
   A high vacuum gate valve for skimmed molecular beam experiments is described.
   It is designed with a very short extent of only 10~mm along the molecular
   beam axis to minimize the distance between the molecular beam source and the
   experiment to provide the maximum molecular flux to the experiment. At the
   same time it provides free space on both sides of the skimmer to not disturb
   the supersonic expansion in front of the skimmer, to give optical access to
   the full distance between beam source and skimmer, and to allow for placing
   electrostatic devices very close behind the skimmer. The gate valve allows to
   maintain high vacuum conditions ($10^{-8}$~mbar) in the experimental chamber
   while the source chamber is brought up to atmospheric pressure for
   modifications or maintenance. The valve can be operated from outside the
   vacuum chamber while maintaining vacuum conditions in all chambers.\\[1em]
   Copyright 2006 American Institute of Physics. This article may be downloaded
   for personal use only. Any other use requires prior permission of the author
   and the American Institute of Physics. The following article appeared in
   \emph{Rev.~Sci.~Instrum.}~\textbf{77}, 016106 (2006) and may be found at
   \url{http://dx.doi.org/10.1063/1.2162456}.
\end{abstract}
\maketitle%

Molecular beams are an indispensable tool in physical chemistry and molecular
physics. They provide extreme cooling from ambient or elevated temperatures down
to 1~K and are used for a large variety of experiments
\cite{_Scoles:MolBeam:1,_Scoles:MolBeam:2}. In most applications the molecular
beam is skimmed a few cm downstream from the nozzle to reduce the gas-load in
the experimental chamber and to select only the most intense part of the beam,
or to collimate the beam for reducing the transverse velocity spread.

Small molecules embedded in a supersonic beam can be focused and state-selected
using electrostatic lenses \cite{Reuss:StateSelection} and they can be
decelerated using switched electric fields as in a Stark decelerator
\cite{Bethlem:IntRevPhysChem22:73}. In such experiments it is desirable to
accept the largest attainable fraction of the skimmed beam. This requires,
besides other design criteria, to place the device as close to the skimmer as
possible, as the particle flux falls off rapidly with distance. At the same time
such devices are operated at high electric fields and must be high-voltage
conditioned whenever they have been exposed to air. Therefore, it is highly
desirable to allow for a vacuum separation of the electrostatic device and the
beam source, where changes to the nozzle for different experiments or regular
maintenance might be required. Commercial gate valves, however, cannot be used
since they take up too much space.

\begin{figure}[b]
   \centering
   \includegraphics{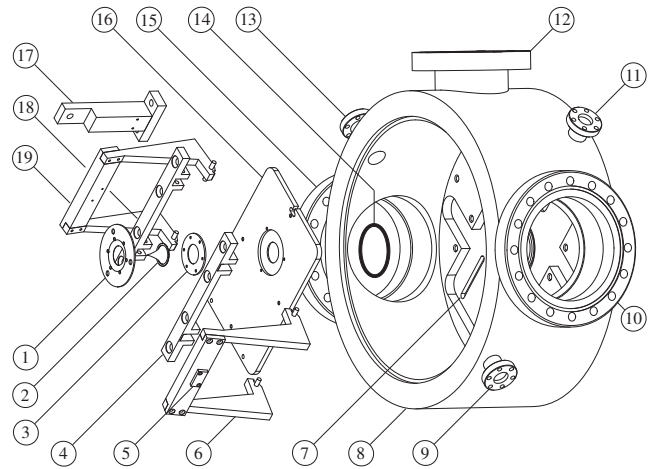}
   \caption{Explosion drawing of the CF~250 front flange and gate valve. The
      individual parts are: (1) upper skimmer mount, (2) skimmer, (3) lower
      skimmer mount, (4,~18) shaft holders, (5) hardened steel plate, (6,~19)
      lever arms, (7) slots for adjustment pins, (8) chamber flange, (9,~11,~13)
      CF~16 flanges, (10,~15) CF~100 flanges, (12) CF~63 flange, (14) Viton
      O-ring, (16) sledge, (17) linear motion feedthrough guide; see text for
      details.}
   \label{fig:explosion}
\end{figure}
In this note we present an extremely thin gate valve design with an extent along
the molecular beam axis of only 10~mm. It allows to keep the experimental
chambers under high vacuum ($10^{-8}\text{~mbar}$) while the source chamber is
vented to atmospheric conditions. The valve is implemented in the source chamber
of a new Alternate Gradient decelerator experiment for the deceleration of large
\mbox{(bio-)}molecules. This experiment needs different complex molecular beam
sources to transfer the molecules to be studied into the gas phase, for example
a laser desorption setup. These sources require regular maintenance, as well as
optical access to the whole distance between nozzle and skimmer. Furthermore, it
is important to provide free space around the skimmer in order to not disturb
the supersonic expansion. At the same time it is desirable to minimize the
distance from the molecular beam source to the decelerator, which is placed
directly behind the skimmer.

Low-profile gate valves have been constructed before, but only a few are
practicable to isolate skimmed molecular beam sources from high vacuum systems
\cite{Chaban:RSI64:2391,Stolow:JVSTA14:2669,Marceca:RSI68:3258}. They all have
in common that they incorporate the skimmer into the valve. Marceca et al.\
\cite{Marceca:RSI68:3258} simply close the skimmer opening from the back with a
needle, which in our experiment would interfere with the decelerator placed
directly behind the skimmer. The design by Stolow \cite{Stolow:JVSTA14:2669} is
a simplified extension of the valve by Chaban et al.\ \cite{Chaban:RSI64:2391}.
Both designs are, however, two and a half times thicker than our design and the
implementation in a separate conflat flange does not allow for simple optical
access directly at the tip of a short skimmer, as is advantageous in our
experiments.

\begin{figure}[b]
   \centering
   \includegraphics{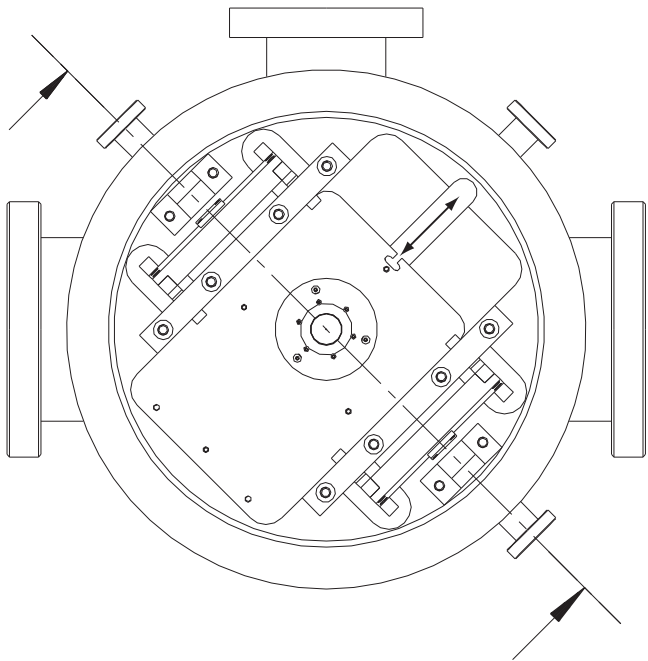}
   \caption{Front view showing the overall layout of the assembled gate valve
      from the position of the beam source. In the center the molecular beam
      skimmer (2) is seen in the skimming position. The line depicting the
      cut-plane for the drawing in figure~\ref{fig:cut} crosses the shaft
      holders (4, 18), the levers (6, 19), the linear motion feedthrough guide
      (17), the flange wall (8), and the CF~16 flanges (9, 13) for fixing the
      linear motion feedthroughs. On the left and right are CF~100 flanges (10,
      15) for optical access and at the top is the flange (12) for LIF detection
      or optical access. At the top right is the CF~16 flange (11) for the
      sliding linear motion feedthrough, whose operating range for moving the
      sledge is depicted by the double-headed arrow.}
   \label{fig:front}
\end{figure}
\begin{figure}[b]
   \centering
   \includegraphics{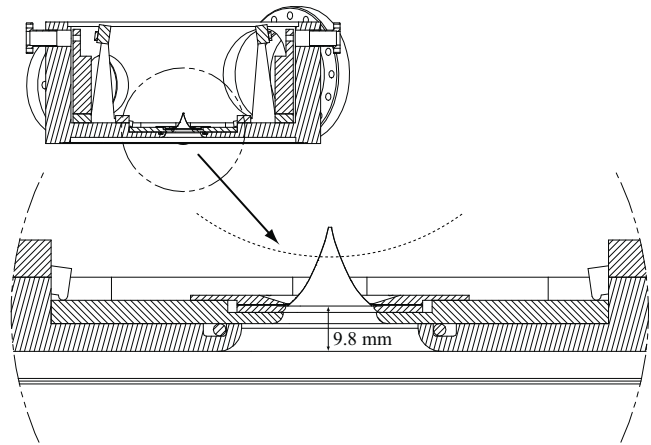}
   \caption{Cut through the plane depicted in figure~\ref{fig:front}. The
      visible parts around the center are (bottom to top) the chamber wall (8),
      the Viton O-ring (14), the sledge (16), the lower skimmer mount (3), the
      skimmer (2), the upper skimmer mount (1), and the higher part of the wall
      (8) in the back. At the sides the shaft holders (4, 18) and the tips of
      the lever arms (6, 19) are visible. The dashed arc shows the open view
      through the CF~100 flanges (10, 15) of the chamber.}
   \label{fig:cut}
\end{figure}
In our design, the thin gate valve is integrated into a homebuilt CF~250 flange,
that incorporates the gate valve in the wall separating source and deceleration
chamber. There are several smaller flanges at right angle to provide optical and
mechanical access to the source chamber. The CF~250 flange is welded on a
commercial CF~250 four-way cross, that is used as the source chamber of our
Alternate Gradient deceleration experiment.

In the explosion drawing of figure~\ref{fig:explosion} the individual parts,
which are described in the following, are shown. Figures~\ref{fig:front} and
\ref{fig:cut} show the arrangement of the parts in the assembled gate valve and
in figure~\ref{fig:cut} the height of the valve assembly is depicted.
The gate valve is embedded in the CF~250 front flange (8) of the source chamber
of our experimental setup. The flange has two CF~100 flanges (10, 15) at right
angles at the left and right for optical and mechanical access to the molecular
beam source (not shown) and the skimmer (2) from the sides, and a CF~63 (12)
flange for access from the top. Integrating these flanges into the CF~250
flange/wall assembly gives optical access to the full distance between the
nozzle and the skimmer. We have made use of this to excite CO molecules to their
metastable $a^3\Pi$-state and to detect laser induced fluorescence (LIF) signal
from metastable CO \cite{Jongma:JCP107:7034} and benzonitrile
\cite{Borst:CPL350:485}. The wall separating source and experimental chamber has
a central hole of 38~mm diameter. This hole is big enough to not disturb a
passing molecular beam and also to provide large enough distances from the
high-voltage electrodes in our decelerator. In a groove milled into the wall on
the source-side a sledge is mounted to provide the valve closure. The vacuum
seal is provided by a $44\text{~mm}\times3.5\text{~mm}$ diameter Viton O-ring
(14) mounted in a circular groove of the wall and pressed between the sledge
(16) and the wall. The sledge itself has a circular groove of 44~mm diameter in
which the skimmer assembly, consisting of an upper (1) and lower (3) skimmer
mount and the skimmer (2) itself, is mounted. The sledge (16) is moved inside
the groove by a linear motion feedthrough (Huntington L-2111) mounted on a CF~16
flange (11) into skimming or sealing position. The traveling distance of the
sledge between the two positions is 49~mm. In figure~\ref{fig:front} the sledge
is shown in skimming position. In this position the skimmer position is adjusted
onto the molecular beam axis of the experiment. When closing and opening the
gate valve the skimmer is reproducibly positioned with a precision of 0.01~mm by
two adjustment pins below the sledge moving inside precisely machined slots (7).

When the sledge is positioned it is pressed against the O-ring in the wall by
two double rocker arm levers (6, 19) positioned symmetrically around the O-ring.
A pair of lever arms is connected by a crossbar at the top to operate two levers
simultaneously. Their axes are mounted close to the chamber wall in shaft
holders (4, 18) which are fixed on the wall. The levers are operated by two
linear motion feedthroughs (Huntington L-2111) mounted on CF~16 flanges (9, 13).
The linear motion feedthroughs are protected against angular forces by guides
assembled close to the lever arm crossbars (17, front guide not shown in
figure~\ref{fig:explosion}, see figure~\ref{fig:front} for its position). To fix
the sledge, the linear motion feedthroughs push against hardened steel plates
(5) on the back of the lever arm crossbar. The levers then push the sledge
symmetrically against the O-ring with a 15:1 transmission ratio. A pressure
difference between the source and deceleration chamber will automatically
provide an additional sealing force on the sledge. When there is no large
pressure difference between the two sides of the valve (i.\,e.\ both chambers at
high vacuum or both chambers vented) and the lever arms are released, the sledge
can be moved perpendicular to the molecular beam axis by a 50.8~mm linear motion
vacuum feedthrough (Huntington L-2111) mounted on a CF~16 flange (11) directly
connected to the sledge.

The described gate valve is implemented in a new vacuum system used for the
Alternate Gradient deceleration \cite{Bethlem:PRL88:133003} of large molecules.
After the original assembly of the chambers we have carefully noted the readings
of all linear motion feedthroughs when the sledge is fixed in operation
conditions, i.\,e.\ when the skimmer is on the beam axis, when the sledge is in
sealing position, and while the sledge is moved from one configuration to the
other, i.\,e.\ when the lever arms are loose. A skimmer (Beam Dynamics, model~1,
1~mm orifice) is mounted on the sledge and carefully aligned onto the molecular
beam axis with the sledge fixed in the corresponding position. Then the chambers
are evacuated and the decelerator is high-voltage conditioned. When the sledge
is moved to sealing position the source chamber can be vented for work on the
molecular beam source. Under these conditions the pressure in the deceleration
chamber stays at $1\times{}10^{-8}\text{~mbar}$, protecting the multi channel
plates from air, and afterward no reconditioning of the high-voltage electrodes
is necessary. This process has been performed numerous times over the last year
and no recognizable degradation is observed.

The space-effective gate valve presented here is of general use for all
molecular beam experiments where the cost of venting the full system is high
compared to venting the source-chamber only. This can be due to air-sensitive
devices, as high-voltage electrodes, sensitive detectors, or also simply due to
vacuum requirements, where venting of the full experiment requires long pumping
times due to large volumes or high vacuum requirements in the experimental
chambers. The presented gate valve is especially useful for applications that
require a short distance between the nozzle and the experimental device, for
example hexapole focusing or Stark deceleration of molecular beams. It is
invaluable when sources are used that require regular maintenance or cleaning,
i.\,e.\ laser desorption or ablation sources, discharge sources, pyrolysis
sources, or sources that are simply heated or cooled to extreme temperatures.

\begin{acknowledgments}
   The authors are grateful to the FHI machine shop for their expert fabrication
   of the described device.
\end{acknowledgments}

\bibliographystyle{apsrev}
\bibliography{string,mp}

\end{document}